\documentclass[conference]{IEEEtran}
\IEEEoverridecommandlockouts
\usepackage{cite}
\usepackage{amsmath,amssymb,amsfonts}
\usepackage{algorithmic}
\usepackage{graphicx}
\usepackage{lipsum}
\usepackage{float}
\usepackage{textcomp}
\usepackage{xcolor}
\usepackage{balance}
\def\BibTeX{{\rm B\kern-.05em{\sc i\kern-.025em b}\kern-.08em
    T\kern-.1667em\lower.7ex\hbox{E}\kern-.125emX}}
\begin{document}

\title{Outage Probability in Network Coding Based Cooperative Wireless Networks over Nakagami-$m$ Fading Channels
}

\author{
\IEEEauthorblockN{Elias Benamira\IEEEauthorrefmark{1}, 
						Fatiha Merazka\IEEEauthorrefmark{1}, Güneş Karabulut Kurt\IEEEauthorrefmark{2} }

\IEEEauthorblockA{\IEEEauthorrefmark{1}LISIC Laboratory, Telecommunications Department.
USTHB University,\\P. O. Box 32 El Alia, Algiers, Algeria, \{ebenamira,fmerazka\}@usthb.dz, http://www.usthb.dz\\
						\IEEEauthorrefmark{2}Department of Communications and Electronics Engineering,\\ Istanbul Technical University,34469, Maslak, Istanbul, Turkey, gkurt@itu.edu.tr, http://www.itu.edu.tr \\
	}
}

\maketitle

\begin{abstract}
In this paper, we develop an accurate closed-form analytic expression of the outage probability for each source-destination (S-D) pair in the two S-D pairs two relays wireless network, where cooperative network coding is applied over Nakagami-$m$ fading channels. For different values of the shape parameter $m$, our analytic relation is validated with Monte-Carlo simulations of the outage probability using the overall equivalent signal-to-noise ratio (SNR) within each source-destination pair. The outage probability is also provided for the extended versions of the network consisting of multiple relays and multiple S-D pairs. Furthermore, we investigate the role of varying the fading factor ($m$ values) on different links in the end-to-end outage probability performance, which yields interesting results that may be crucial for relay selection and power allocation procedures. Moreover, we derive the diversity order for the generalized extended version of cooperative network coded wireless networks.
\end{abstract}

\begin{IEEEkeywords}
Cooperative communication, multiple S-D pairs, Nakagami-$m$ fading channel, network coding, outage probability. 
\end{IEEEkeywords}

\section{Introduction}
Developing highly reliable wireless networks is a challenge that compels researchers to explore novel strategies combining network information theory and signal processing for telecommunications principles for a better optimization of wireless networks resource management and a higher efficiency accomplishment. Cooperation and network coding in wired and wireless networks has, since its proposal and first applications \cite{1}, been proven to provide significant improvement to the network performance in terms of efficiency and security. Besides the considerable throughput gain achieved \cite{2,3}, which is one of the main important targets of 5G technology wireless networks, cooperative network coding increases the diversity order and security of data transmission over the network \cite{4,5} while reducing the overall system power consumption.

In network coded cooperation (NCC) paradigm, the relay in charge of network coding (NC) function performs a linear combination of the $N$ symbols received from $N$ source nodes and broadcasts the resulting symbol to different destinations or other relays. Hence, the number of transmissions over the network is reduced which increases the throughput gain. Furthermore, this process enhances the security of each message $m_k$ being transmitted considering that all other messages $m_i (i=1, ..., N, i\neq k)$ contribute to ensure its secrecy \cite{6}. The use of exclusive OR (XOR) operator is straightforward and practical for network coding application. The NC resulting symbol at relay $R$ is given by $m_{NC}=m_R=\bigoplus_{i=1}^{N} m_i$ where $\bigoplus$ represents the summation symbol over Galois field $GF(M)$. Each destination will then recover its destined message by performing the same XOR operation on its received signals. Cooperative communication, combined with network coding technique, has the key benefits of combating the fading inherent to wireless channels and increasing capacity and diversity gains \cite{7}.

In wireless environments, transmission of signals is subject to the effects of various phenomena such as reflection, refraction, diffraction, absorption and scattering which results in what we call multi-path received signal. In radio channels, these effects are called flat multi-path fading and create small scale fading alterations. These alterations are usually modeled by Rice, Rayleigh and Nakagami-$m$ channel fading models. Rice and Rayleigh distributions describe multi-path effects and Nakagami-$m$ is a
general form that can replace them both.\newline
Several recent publications have suggested the use of Nakagami-$m$ and Weibull fading models to provide a generalized description of fading in wireless systems. One of the most versatile fading model is described by the Nakagami-$m$ distribution. It has a greater flexibility and accuracy in matching some experimental data than Rayleigh, log-normal, Weibull or Rician distributions. The Nakagami-$m$ fading distribution was first used for modeling ionospheric and tropospheric fast fading channels and has been widely adopted for multi-path modeling in wireless communications. It provided the best fit to some urban multi-path data \cite{8}.\newline
Many previous works have been treating network coding and cooperation application on different wireless networks and few of them investigated and derived the outage probability of the end-to-end transmission over a NCC wireless network. In \cite{9}, authors considered NC-based cooperative systems where multiple sources are communicating with one destination via multiple relays. Their system, based on code division multiple access (CDMA) transmission technique, showed better performance over traditional cooperative systems by reducing the outage probability. A two-way decode-and-forward (DF) non-orthogonal multiple access (NOMA) scheme is proposed in \cite{10} where the relay performs digital network coding (DNC) in combination with NOMA. The authors derived closed-form theoretical expression of the outage probability over Rayleigh fading channels. Their proposed DNC based two-way NOMA (TWNOMA) protocol outperforms the conventional DNC protocol when the relay is located in some optimal positions between both sources. Authors in \cite{11} derived theoretical analytic expressions of outage probabilities for decode-and-forward, selective DF and cooperative network coding (CNC) protocols in multiple relay systems over Rayleigh fading channels. After verification of analytic expressions of outage probabilities by simulations, they concluded the CNC provides the best performance in increasing system throughput and diversity gain of the relay-based cooperative system. Communication performance in vehicular ad hoc network (VANET) is investigated in \cite{12}. Authors assumed vehicles equipped with multiple antennas and applied non-binary network coding scheme to prove that multiple input multiple output (MIMO) based schemes give better performance over non-binary NC (NBNC) and generalized dynamic NC (GDNC) schemes over Rayleigh fading model channels. Outage behavior of a system composed of one source, multiple relays and one destination under Nakagami-$m$ fading environment was investigated in \cite{13}. Using instantaneous power-based approach, the authors derived the exact closed-form expression of outage probability and found the optimal number of relays required for a desired outage performance. Analysis of symbol error rate (SER) performance of space-time network coding (STNC) over independent but not necessarily identically distributed (i.n.i.d.) Nakagami-$m$ fading channels is conducted in \cite{14}. Authors assumed multiple sources communicating with one destination via multiple amplify and forward (AF) relays. They derived the exact expression of overall end-to-end received signal-to-noise ratio (SNR) and its moment generating function (MGF). A closed-form expression of the SER with $M$-PSK and $M$-QAM modulations is also presented in this paper. Their simulations illustrated that the diversity order of STNC with multiple AF relays is a sum function of the fading index of the direct link and the minimal fading indices of the multiple two-hop links. Namdar \textit{et al.} \cite{15} analysed the outage probability performance in the high SNR region along with the average bit error rate (BER) performance using pulse position modulation (PPM) with timing error rate for hybrid radio frequency/visible light communication (RF/VLC) links in indoor relay-assisted systems. They implemented differential evolution algorithm (DEA) to determine the optimal PDF approximation value achieving the best outage probability performance.\newline
An energy-constrained cooperative network with application of NOMA over Nakagami-$m$ fading channels is studied in \cite{16}. In their system model composed of one base station, one relay and multiple users (destinations), authors assumed the relay equipped with energy harvesting devices. They derived a closed-form expression for the outage probability at each end user along with the upper and lower bounds. Using analytic and simulated results, they demonstrated that NOMA application yields significant improvement in spectral efficiency in comparison to orthogonal multiple access (OMA) strategy. Authors in \cite{17} derived the exact outage probability expression of their proposed signal space cooperative (SSC) scheme. Furthermore, the asymptotic approximation of their closed-form outage probability shows a clear gain in diversity and coding gain. Multiple relay selection problem in cooperative communications is studied in \cite{18}. Authors derived a tractable bound for the outage probability constraint where the multiple relay selection problem is formulated as a mixed integer optimization problem with the on-and-off power mode at the relays.

Our paper offers the following main contributions:
We derive the exact closed-form expression of the outage probability in two S-D pairs two relays wireless network over Nakagami-$m$ fading channels. The analytic expression obtained is validated by the simulated version that uses the equivalent instantaneous SNR over the end-to-end link between each source-destination pair. Furthermore, we give an extended version of the outage probability for the general extensions of the considered wireless network topology with multiple relays and multiple source-destination pairs. Besides, we investigate how Nakagami $m$ factor variation over the different links affects the outage probability behavior. \newline
The rest of this paper is organized as follows:
In Section 2, we present a description of how the information is handled in our system model and the channel fading model. Section 3 contains the detailed developments of outage probability expressions investigated in this paper. The simulations carried out along with the results obtained are provided in Section 4. A conclusion of our work is finally given in Section 5. 

\section{System Model}

\begin{figure}[!t]
\centerline{\includegraphics[scale=0.4]{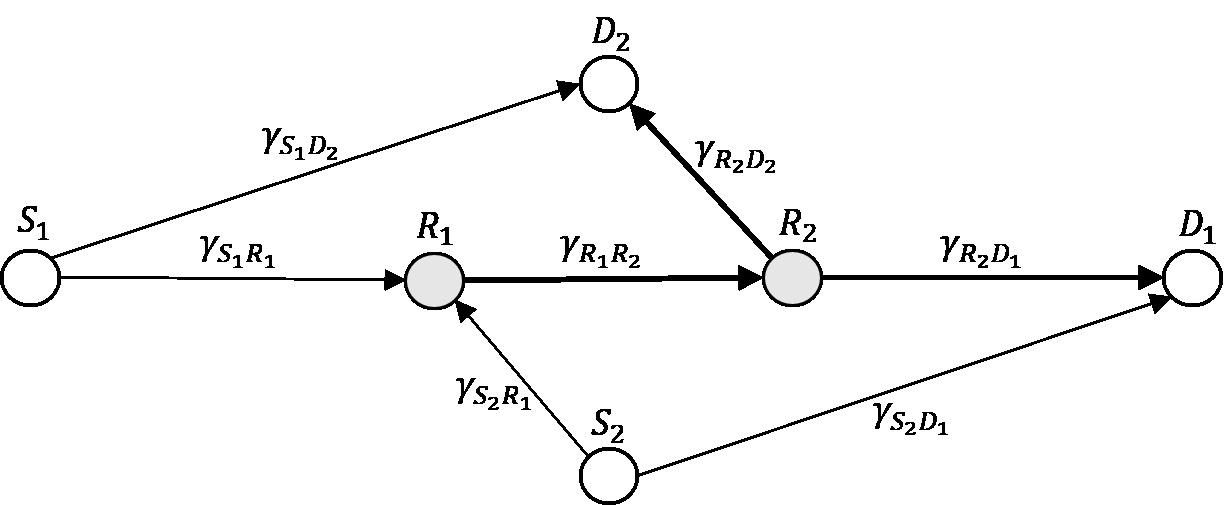}}
\caption{Considered system: Two S-D pairs two relays wireless network}
\label{fig_1}
\end{figure}

The system studied in this paper is the two S-D pairs two relays wireless network illustrated in Fig. \ref{fig_1} where two source-destination pairs are communicating via two relays in series. In such network, each source doesn't have direct radio access to its destination but has a direct link with the other destination.
The first relay $R_1$, working in half-duplex mode, receives in time-slot $T_1$ message $x_1$ from source node $S_1$ and then $x_2$ from $S_2$ in time-slot $T_2$ in a time division multiple access (TDMA) transmission mode. We consider both sources are transmitting binary symbols from the finite field $\mathbb{F}_2$ and hence we choose binary phase-shift keying (BPSK) digital modulation technique for transmission.\newline

The general form of received signal over each link $(i \rightarrow j)$, is given by

\begin{equation}\label{1.1}
y_{ij}=\sqrt{P_i}h_{ij}x_i+n_{ij}
\end{equation}
where $i\in \{S_1, S_2, R_1, R_2\}, j\in \{R_1, R_2, D_1, D_2\}$, $P_i$ is the transmit power of node $i$, $h_{ij}$ is the channel fading coefficient of transmission from node $i$ to node $j$ and $n_{ij}$ represents the complex additive white Gaussian noise (AWGN) with zero mean and variance $\sigma_{n_{ij}}^2$ at receiver $j$, $n_{ij} \sim \mathcal{C}\mathcal{N}(0, \sigma_{n_{ij}}^2)$.\newline

We, first, define the following applications \cite{19}:\newline
1.~~$\mathcal{M}_2 : \mathbb{F}_2 \rightarrow \mathbb{C}$ that maps each symbol $m_i \in \mathbb{F}_2$ into the corresponding BPSK constellation $x_i$, i.e. $x_i = \mathcal{M}_2(m_i)$.\newline
2.~~$\mathcal{E} : \mathbb{C} \rightarrow \mathbb{C}$  that estimates the received version $\tilde{x}_{ij}$ of the original one $x_i$ at the receiver $j$, i.e. $\tilde{x}_{ij}=\mathcal{E}(y_{ij})=\mathcal{E}(\sqrt{P_i}h_{ij}x_i+n_{ij})=(\sqrt{P_i}h_{ij}x_i+n_{ij})h_{ij}^*$, where $h_{ij}^*$ denotes the complex conjugate of $h_{ij}$. \newline In this paper, we assume that all transmitting nodes have the same transmit power $P_i=P$ and perfect channel state information (CSI) available at each receiver node.\newline 
The relay $R_1$ recovers both messages $\tilde{m}_{S_1R}=\mathcal{M}_2^{-1}(\tilde{x}_{S_1R_1})=\mathcal{M}_2^{-1}(\mathcal{E}(y_{S_1R_1}))$ and $\tilde{m}_{S_2R_1}=\mathcal{M}_2^{-1}(\tilde{x}_{S_2R_1})=\mathcal{M}_2^{-1}(\mathcal{E}(y_{S_2R_1}))$ before applying network coding operation XOR on them to generate $m_{R_1}=m_{NC}=\tilde{m}_{S_1R_1} \oplus \tilde{m}_{S_2R_1}$ that it modulates and transmits to relay $R_2$. The network coded (NC) symbol is then amplified at $R_2$ (working in AF mode) with amplification factor $\beta_{R_1R_2}=\sqrt{\frac{P_{R_2}}{P_{R_1}|h_{R_1R_2}|^2 + \sigma_{R_1R_2}^2}}$ and forwarded (broadcast) to destination nodes $D_1$ and $D_2$. Destinations $D_1$ and $D_2$ recover their respective messages by applying the same operation on their received signals, i.e. 
\begin{equation}\label{1}
\tilde{m}_1=\tilde{m}_{R_2D_1} \oplus \tilde{m}_{S_2D_1}= \mathcal{M}_2^{-1}(\mathcal{E}(y_{R_2D_1})) \oplus \mathcal{M}_2^{-1}(\mathcal{E}(y_{S_2D_1}))
\end{equation}
and 
\begin{equation}\label{2}
\tilde{m}_2=\tilde{m}_{R_2D_2} \oplus \tilde{m}_{S_1D_2}= \mathcal{M}_2^{-1}(\mathcal{E}(y_{R_2D_2})) \oplus \mathcal{M}_2^{-1}(\mathcal{E}(y_{S_1D_2}))\nonumber
\end{equation}
respectively.\newline

All links in our network are subject to Nakagami-$m$ model fading channel with the probability density function (PDF) \cite{20}
\begin{equation}\label{3}
f(\mathcal{X})=\frac{2m^m\mathcal{X}^{2m-1}}{\Gamma(m)\mu^m}e^{\frac{-m\mathcal{X}^2}{\mu}}
\end{equation}
where 
\begin{equation}
\mu = E(\mathcal{X}^2) \text{ and }
m=\frac{\mu^2}{(\mathcal{X}^2-\mu)^2}\geq\frac{1}{2}\nonumber
\end{equation}  
represents the shape factor of Nakagami-$m$ distribution and $\Gamma(.)$ is the Gamma function defined by 
\begin{equation}
\Gamma(x)=\int_{0}^{\infty} a^{x-1} e^{-a} da\nonumber
\end{equation} 
\newline
In wireless networks links, $\mathcal{X}^2=|h_{ij}|^2$.

Let $\gamma_{ij}=\frac{|h_{ij}|^2P_i}{\sigma_{ij}^2}$ be the instantaneous SNR over the link $(i \rightarrow j)$, then (\ref{3}) becomes \cite{21}
\begin{equation}\label{4}
f(\gamma_{ij})=\frac{m_{ij}^{m_{ij}}\gamma_{ij}^{m_{ij}-1}}{\bar{\gamma}_{ij}^{m_{ij}}\Gamma(m_{ij})}e^{\frac{-m_{ij}\gamma_{ij}}{\bar{\gamma}_{ij}}}
\end{equation}
where $\bar{\gamma}_{ij}=E(\gamma_{ij})$ is the average (mathematical expectation) value and $m_{ij}$ is the shape factor of link $(i\rightarrow j)$.\newline
For the correspondence with Rice distribution $m=\frac{(K+1)^2}{2K+1}$, $K$ being the Rician fading factor.
The special case of Nakagami shape factor $m=1$ represents Rayleigh fading model.\newline

\section{Outage Analysis}
To derive the outage probability of the wireless network in question subject to NCC application, we consider the source-destination pair $(S_1 - D_1)$ where the participating links are $(S_1\rightarrow R_1), (R_1\rightarrow R_2), (R_2\rightarrow D_1), (S_2\rightarrow R_1)$ and $(S_2\rightarrow D_1)$.\newline
To the five links that intervene in message transmission from source node $S_1$ to destination node $D_1$ correspond five outage events $E_{S_1R_1}$, $E_{R_1R_2}$, $E_{R_2D_1}$, $E_{S_2R_1}$ and $E_{S_2D_1}$ respectively. The resultant outage event $E_{S_1D_1}$ of the end-to-end link $(S_1 \rightarrow D_1)$ has the following expression
\begin{equation}\label{5}
E_{S_1D_1}=E_{S_2D_1} \cap \bigg(E_{R_2D_1}\cup E_{R_1R_2} \cup \Big(E_{S_1R_1} \cap E_{S_2R_1}\Big)\bigg)
\end{equation}
The probability of event $E_{S_1D_1}$ represents the NCC two S-D pairs two relays network outage probability, i.e.
\begin{align}\label{6}
P_{out} &=P_r\big\{E_{S_1D_1}\big\}\nonumber\\
        &=P_r\Big\{E_{S_2D_1} \cap \Big(E_{R_2D_1}\cup E_{R_1R_2} \cup \Big(E_{S_1R_1} \cap E_{S_2R_1}\Big)\Big)\Big\}
\end{align}
With the assumption of mutually independent links, (\ref{6}) can be written as
\begin{align}\label{7}
P_{out}&=P_r\big\{E_{S_2D_1}\big\}  \Big(P_r\big\{E_{R_2D_1}\big\}+P_r\big\{E_{R_1R_2}\big\}\nonumber\\
       &+P_r\big\{E_{S_1R_1}\big\} P_r\big\{E_{S_2R_1}\big\}\Big)
\end{align}
or simply
\begin{equation}\label{8}
P_{out}=P_{out}^{S_2D_1} \Big(P_{out}^{R_2D_1}+P_{out}^{R_1R_2}+P_{out}^{S_1R_1}  P_{out}^{S_2R_1}\Big)
\end{equation}
where $P_{out}^{ij}$ designates the outage probability of each link $(i\rightarrow j)$.\newline

The mutual information $I_{ij}$ between node $i$ and node $j$ is given by \cite{22}
\begin{equation}\label{9}
I_{ij} = log_2(1+\gamma_{ij})
\end{equation}
An outage occurs on link $(i\rightarrow j)$ if $I_{ij}$ is lower than a target information rate (spectral efficiency) $R_t$ in bits per second per Hertz (bps/Hz), i.e.
\begin{equation}\label{10}
log_2(1+\gamma_{ij})<R_t
\end{equation}
or, equivalently, if
\begin{equation}\label{11}
\gamma_{ij}<2^{R_t}-1
\end{equation}

In other words, outage is the scenario when the instantaneous SNR, due to bad channel realization, is unable to support the desired information rate \cite{23}.

In consequence, we can express the outage probability on link $(i\rightarrow j)$ by
\begin{equation}\label{12}
P_{out}^{ij}=P_r\big\{\gamma_{ij}<\gamma_{th}=2^{R_t}-1\big\}
\end{equation}
where $\gamma_{th}$ is the lowest value of the instantaneous SNR (threshold) that allows reliable transmission of information between node $i$ and node $j$.\newline
From (\ref{4}) and (\ref{12}), we have
\begin{align}\label{13}
P_{out}^{ij}&=F(\gamma_{th})=\int_{0}^{\gamma_{th}}f(\gamma_{ij}) d\gamma_{ij}\nonumber\\
&=\int_{0}^{\gamma_{th}} \frac{m_{ij}^{m_{ij}}\gamma_{ij}^{m_{ij}-1}}{\bar{\gamma}_{ij}^{m_{ij}}\Gamma(m_{ij})}e^{\frac{-m_{ij}\gamma_{ij}}{\bar{\gamma}_{ij}}} d\gamma_{ij}\nonumber\\
&=\Bigg[ \frac{-\Gamma_{inc}\Big(m_{ij}, \frac{m_{ij}\gamma_{ij}}{\bar{\gamma}{ij}}\Big)}{\Gamma(m_{ij})}\Bigg]_0^{\gamma_{th}}\nonumber\\
&=1-\frac{\Gamma_{inc}\Big(m_{ij}, \frac{m_{ij}\gamma_{th}}{\bar{\gamma}{ij}}\Big)}{\Gamma(m_{ij})}
\end{align}
where $F(\gamma_{th})$ is the cumulative density function (CDF) and $\Gamma_{inc}(.,.)$ is the upper incomplete Gamma function defined by
\begin{equation}
\Gamma_{inc}(x,y)=\int_{y}^{\infty} a^{x-1} e^{-a} da\nonumber
\end{equation}

Substituting the probability of outage on each link by expression (\ref{13}) in relation (\ref{8}), we obtain the general closed-form analytic expression of the outage probability for our NCC based wireless network
\begin{align}\label{14}
P_{out}&=\Bigg(1-\frac{\Gamma_{inc}\Big(m_{S_2D_1}, \frac{m_{S_2D_1}\gamma_{th}}{\bar{\gamma}_{S_2D_1}}\Big)}{\Gamma(m_{S_2D_1})}\Bigg)\nonumber\\
&\times\Bigg[ \Bigg(1-\frac{\Gamma_{inc}\Big(m_{R_2D_1}, \frac{m_{R_2D_1}\gamma_{th}}{\bar{\gamma}_{R_2D_1}}\Big)}{\Gamma(m_{R_2D_1})}\Bigg)\nonumber\\
&+\Bigg(1-\frac{\Gamma_{inc}\Big(m_{R_1R_2}, \frac{m_{R_1R_2}\gamma_{th}}{\bar{\gamma}_{R_1R_2}}\Big)}{\Gamma(m_{R_1R_2})}\Bigg)\nonumber\\
&+\Bigg(1-\frac{\Gamma_{inc}\Big(m_{S_1R_1}, \frac{m_{S_1R_1}\gamma_{th}}{\bar{\gamma}_{S_1R_1}}\Big)}{\Gamma(m_{S_1R_1})}\Bigg)\nonumber\\
&\times\Bigg(1-\frac{\Gamma_{inc}\Big(m_{S_2R_1}, \frac{m_{S_2R_1}\gamma_{th}}{\bar{\gamma}_{S_2R_1}}\Big)}{\Gamma(m_{S_2R_1})}\Bigg)\Bigg]
\end{align}

Replacing $\gamma_{th}$ by its expression from (\ref{12}) we have

\begin{align}\label{15}
P_{out}&=\Bigg(1-\frac{\Gamma_{inc}\Big(m_{S_2D_1}, \frac{m_{S_2D_1}(2^{R_1}-1)}{\bar{\gamma}_{S_2D_1}}\Big)}{\Gamma(m_{S_2D_1})}\Bigg)\nonumber\\
&\times\Bigg[ \Bigg(1-\frac{\Gamma_{inc}\Big(m_{R_2D_1}, \frac{m_{R_2D_1}(2^{R_1}-1)}{\bar{\gamma}_{R_2D_1}}\Big)}{\Gamma(m_{R_2D_1})}\Bigg)\nonumber\\
&+\Bigg(1-\frac{\Gamma_{inc}\Big(m_{R_1R_2}, \frac{m_{R_1R_2}(2^{R_1}-1)}{\bar{\gamma}_{R_1R_2}}\Big)}{\Gamma(m_{R_1R_2})}\Bigg)\nonumber\\
&+\Bigg(1-\frac{\Gamma_{inc}\Big(m_{S_1R_1}, \frac{m_{S_1R_1}(2^{R_1}-1)}{\bar{\gamma}_{S_1R_1}}\Big)}{\Gamma(m_{S_1R_1})}\Bigg)\nonumber\\
&\times\Bigg(1-\frac{\Gamma_{inc}\Big(m_{S_2R_1}, \frac{m_{S_2R_1}(2^{R_1}-1)}{\bar{\gamma}_{S_2R_1}}\Big)}{\Gamma(m_{S_2R_1})}\Bigg)\Bigg]
\end{align}

Finally, if we assume all links independent and identically distributed  (i.i.d.) affected by the same Nakagami fading factor $m$, i.e.  $\bar{\gamma}_{S_2D_1}=\bar{\gamma}_{R_1R_2}=\bar{\gamma}_{R_2D_1}=\bar{\gamma}_{S_1R_1}=\bar{\gamma}_{S_2R_1}=\bar{\gamma}$ and $m_{S_2D_1}=m_{R_1R_2}=m_{R_2D_1}=m_{S_1R_1}=m_{S_2R_1}=m$, then (\ref{15}) reduces to

\begin{align}\label{16}
P_{out}&=\Bigg(1-\frac{\Gamma_{inc}\Big(m, \frac{m(2^{R_t}-1)}{\bar{\gamma}}\Big)}{\Gamma(m)}\Bigg)^2\nonumber\\
&\times\Bigg[2+\Bigg(1-\frac{\Gamma_{inc}\Big(m, \frac{m(2^{R_t}-1)}{\bar{\gamma}}\Big)}{\Gamma(m)}\Bigg)\Bigg]\nonumber\\
&=\Bigg(1-\frac{\Gamma_{inc}\Big(m, \frac{m(2^{R_t}-1)}{\bar{\gamma}}\Big)}{\Gamma(m)}\Bigg)^2\nonumber\\
&\times\Bigg(3-\frac{\Gamma_{inc}\Big(m, \frac{m(2^{R_t}-1)}{\bar{\gamma}}\Big)}{\Gamma(m)}\Bigg)
\end{align}

If we consider the common probability\newline $p=1-\frac{\Gamma_{inc}\Big(m, \frac{m(2^{R_t}-1)}{\bar{\gamma}}\Big)}{\Gamma(m)}$, expression (\ref{16}) yields 
\begin{equation}\label{17}
P_{out}=p^2(2+p)=2p^2+p^3\sim p^2 \sim O\left(\frac{1}{SNR}\right)^2
\end{equation}
The diversity order achieved in high SNR values for this network is deduced from (\ref{17}) and equals 2. Actually, this result is valid only for $m = 1$, in the next section we will see how diversity order depends on $m$ value.\newline

The previously derived outage probability expressions can be generalized to more than two source-destination pairs and more than two relays in series.

\begin{figure}[!t]
\centerline{\includegraphics[scale=0.35]{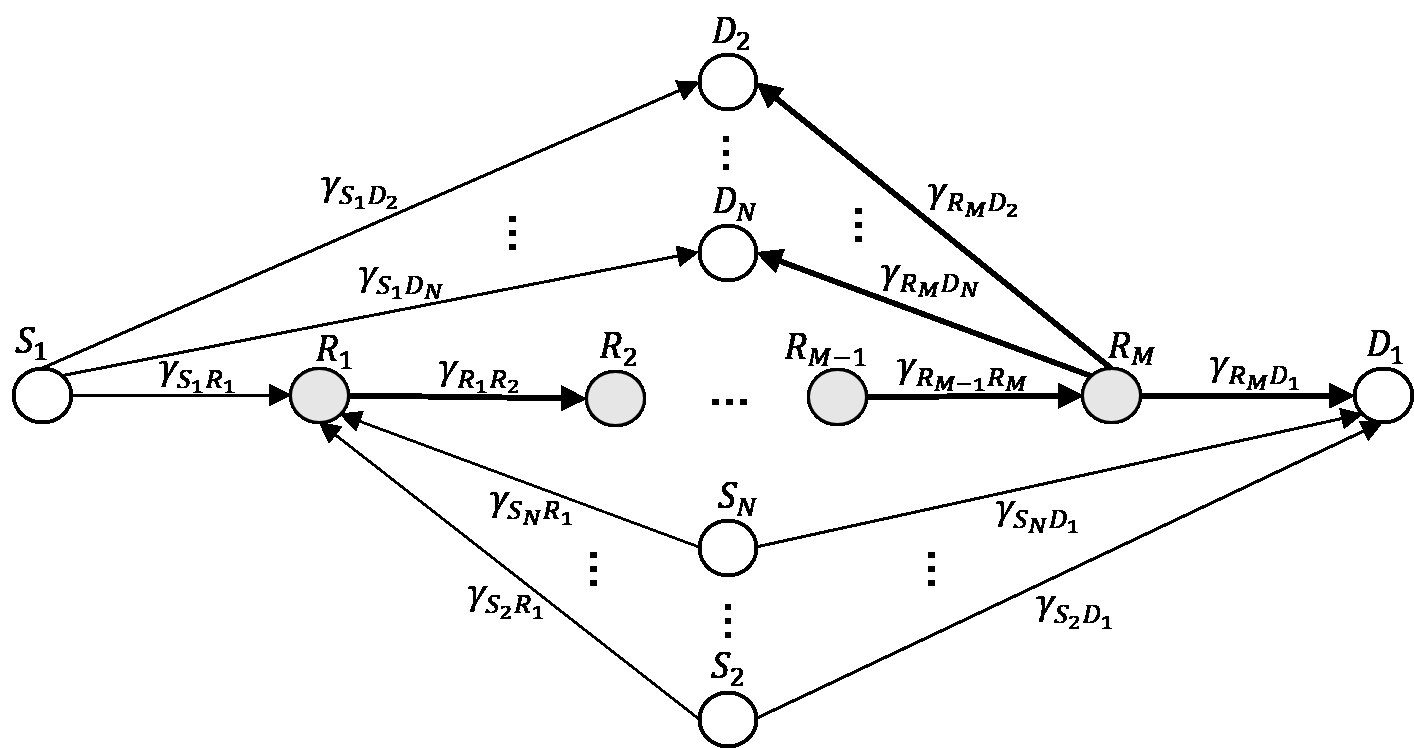}}
\caption{Extended topology of the wireless network with $M$ relays and $N$ source-destination pairs}
\label{fig_2}
\end{figure}

\begin{figure*}[!t]
\begin{equation}\label{18}
\gamma_{S_1D1}^{eq}=
\frac{P_{S_2}|h_{S_2D_1}|^2+P_{R_1}P_{R_2}|h_{R_1R_2}|^2|h_{R_2D_1}|^2\Big(P_{S_1}|h_{S_1R_1}|^2+P_{S_2}|h_{S_2R_1}|^2\Big)}{E\{n_{S_2D_1}^2\}+E\{n_{R_2D_1}^2\}+P_{R_2}|h_{R_2D_1}|^2E\{n_{R_1R_2}^2\}+P_{R_1}P_{R_2}|h_{R_1R_2}|^2|h_{R_2D_1}|^2\Big(E\{n_{S_1R_1}^2\}+E\{n_{S_2R_1}^2\}\Big)}
\end{equation}
\end{figure*}

\begin{figure*}[!t]
\begin{equation}\label{19}
P_{out}^{sim}=
P_r\Bigg\{\frac{P_{S_2}|h_{S_2D_1}|^2+P_{R_1}P_{R_2}|h_{R_1R_2}|^2|h_{R_2D_1}|^2\Big(P_{S_1}|h_{S_1R_1}|^2+P_{S_2}|h_{S_2R_1}|^2\Big)}{E\{n_{S_2D_1}^2\}+E\{n_{R_2D_1}^2\}+P_{R_2}|h_{R_2D_1}|^2E\{n_{R_1R_2}^2\}+P_{R_1}P_{R_2}|h_{R_1R_2}|^2|h_{R_2D_1}|^2\Big(E\{n_{S_1R_1}^2\}+E\{n_{S_2R_1}^2\}\Big)}< \gamma_{th}\Bigg\}
\end{equation}
\end{figure*}

\begin{figure*}[!t]
\begin{equation}\label{20}
P_{out}^{sim}=
P_r\Bigg\{\frac{P_{S_2}|h_{S_2D_1}|^2+P_{R_1}P_{R_2}|h_{R_1R_2}|^2|h_{R_2D_1}|^2\Big(P_{S_1}|h_{S_1R_1}|^2+P_{S_2}|h_{S_2R_1}|^2\Big)}{E\{n_{S_2D_1}^2\}+E\{n_{R_2D_1}^2\}+P_{R_2}|h_{R_2D_1}|^2E\{n_{R_1R_2}^2\}+P_{R_1}P_{R_2}|h_{R_1R_2}|^2|h_{R_2D_1}|^2\Big(E\{n_{S_1R_1}^2\}+E\{n_{S_2R_1}^2\}\Big)} < 2^{R_t}-1\Bigg\}
\end{equation}
\end{figure*}

\begin{figure}[t!]
\centerline{\includegraphics[scale=0.45]{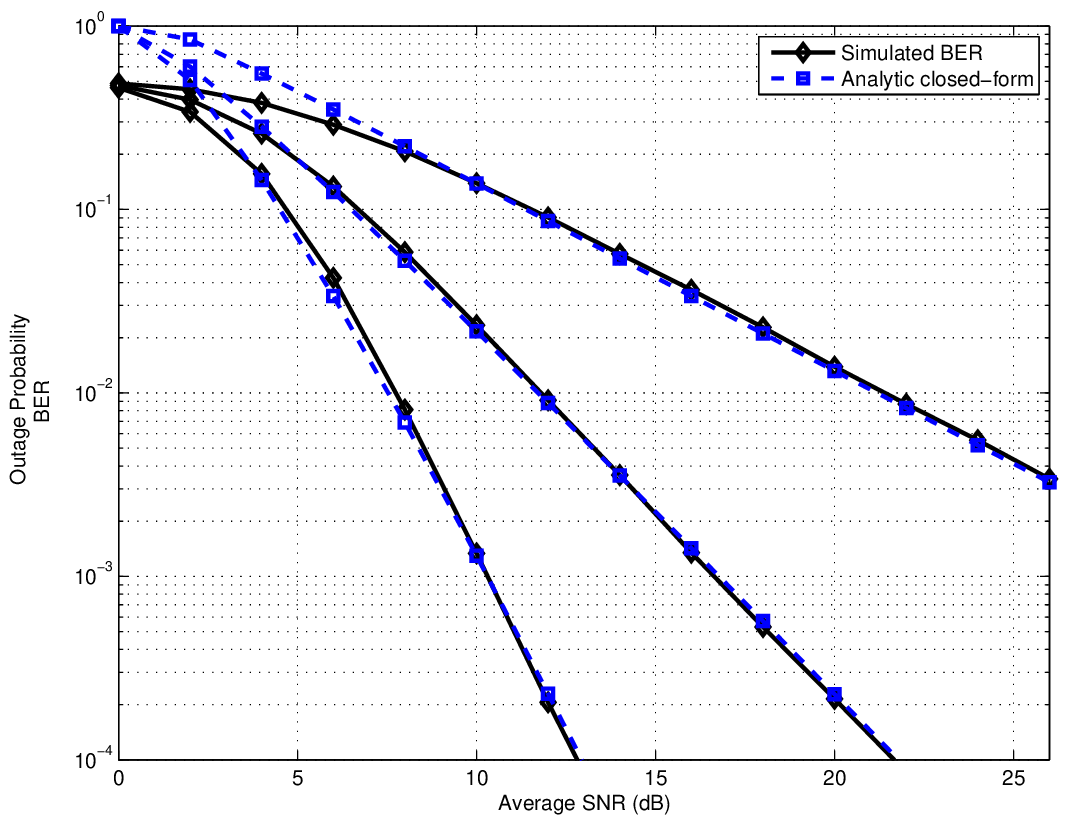}}
\caption{Decreasing analytic outage probability of (\ref{16}) tightly with MC simulated BER for increasing values of $m$ (0.5, 1, 2) and $R_t\approx 1.1~bps/Hz$.}
\label{fig_3}
\end{figure}

\begin{figure}[t!]
\centerline{\includegraphics[scale=0.45]{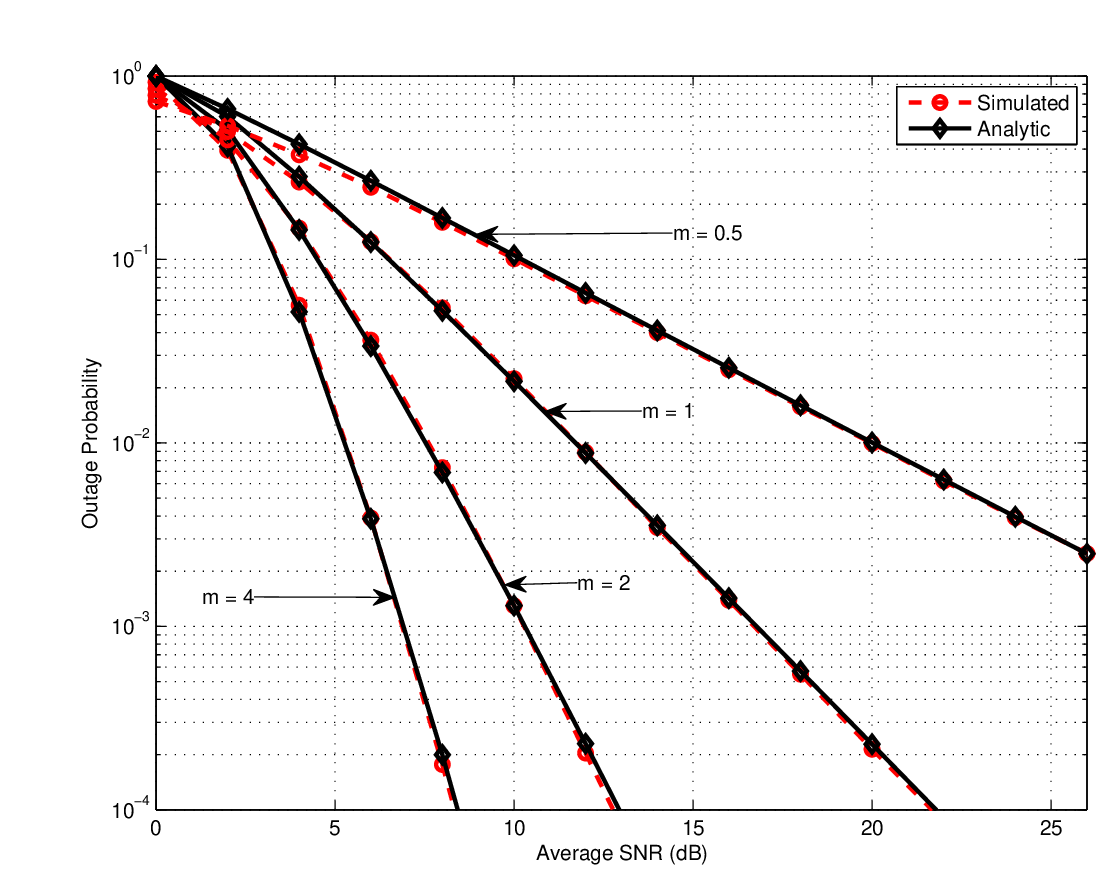}}
\caption{Precise match of simulated with analytic decreasing outage probabilities with increasing values of $m$ ($0.5, 1, 2, 4$).}
\label{fig_4}
\end{figure}

\begin{figure}[t!]
\centerline{\includegraphics[scale=0.45]{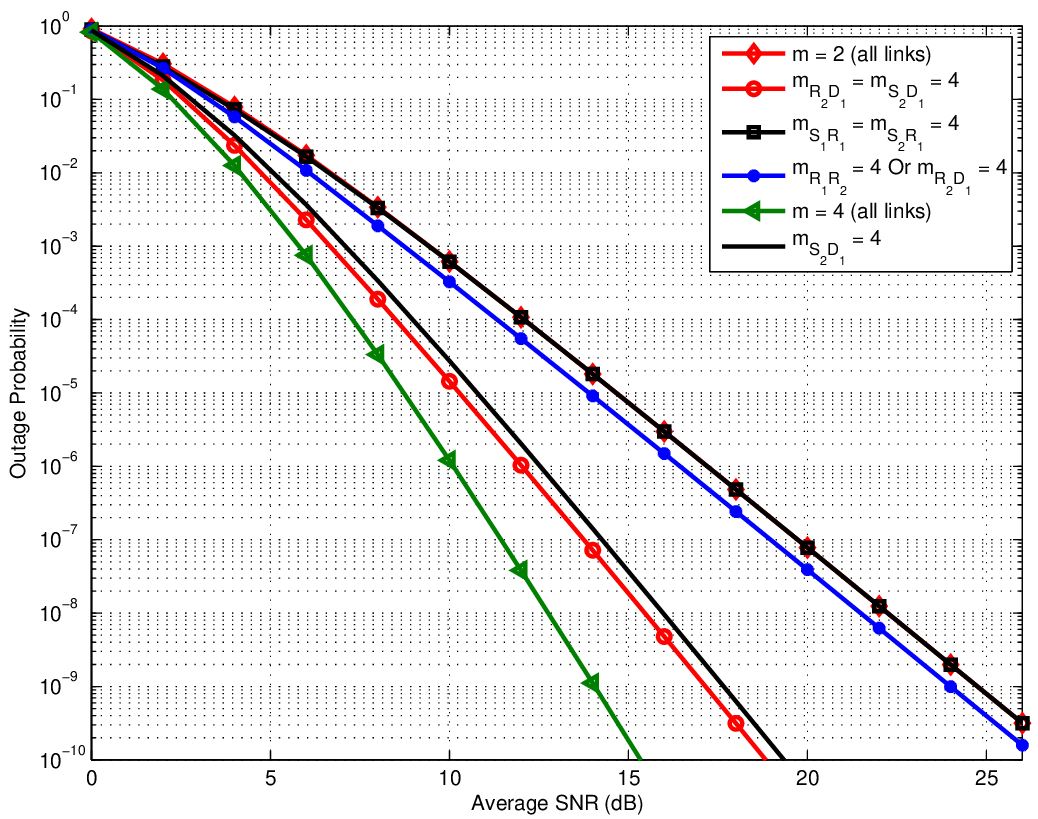}}
\caption{Outage probability of expression (\ref{15}) for $m = 2$ or $m = 4$ on the links and $R_t = 1~bps/Hz$.}
\label{fig_5}
\end{figure}

\newtheorem{theorem}{Theorem}
\begin{theorem}\label{th_1}
\emph{(Extended multiple S-D pairs multiple relays network)}\newline
In an extended version of the network constituted of $N$ source-destination pairs and two relays with $i.i.d.$ Nakagami-$m$ fading channel links and the same value of fading factor $m$, the outage probability expression is given by
\begin{align}\label{21}
P_{out}=&\Bigg(1-\frac{\Gamma_{inc}\Big(m, \frac{m(2^{R_t}-1)}{\bar{\gamma}}\Big)}{\Gamma(m)}\Bigg)^N\nonumber\\
&\times\Bigg[2+\Bigg(1-\frac{\Gamma_{inc}\Big(m, \frac{m(2^{R_t}-1)}{\bar{\gamma}}\Big)}{\Gamma(m)}\Bigg)^{N-1}\Bigg]
\end{align}
Moreover, if we have $M$ ($M>2$) relays $R_1, R_2,...,R_M$ in series instead of two ($R_1$ being always the one in charge of network coding operation and the rest working in AF mode) as illustrated in Fig. \ref{fig_2}, the outage probability of any source-destination pair is then given by
\begin{align}\label{22}
P_{out}=&\Bigg(1-\frac{\Gamma_{inc}\Big(m, \frac{m(2^{R_t}-1)}{\bar{\gamma}}\Big)}{\Gamma(m)}\Bigg)^N\nonumber\\
&\times\Bigg[M+\Bigg(1-\frac{\Gamma_{inc}\Big(m, \frac{m(2^{R_t}-1)}{\bar{\gamma}}\Big)}{\Gamma(m)}\Bigg)^{N-1}\Bigg]
\end{align}
The diversity order achieved in high SNR region, using the result of (\ref{17}), is obtained from
\begin{align}\label{23}
P_{out}&=p^N(M+p^{N-1})\nonumber\\
&=Mp^N+p^{2N-1}\sim p^N \sim O\left(\frac{1}{SNR}\right)^N
\end{align}
and is equal to $N$.

\end{theorem}

In Theorem \ref{th_1}, we assume all links sharing the same Nakagami fading parameter $m$. However, in real life wireless systems, this assumption is seldom true and we have to deal with the generalized version of expression (\ref{24}) that we present in the following theorem.

\begin{theorem}\label{th_2}
\emph{(Generalized multiple S-D pairs multiple relays network extensions)}\newline
Using the same assumptions of Theorem 1 with the exception of distinct value of fading factor $m$ on each link, the outage probability expression has the following general form:

\begin{align}\label{24}
P_{out}=&\prod_{i=2}^{N} \Bigg(1-\frac{\Gamma_{inc}\Big(m_{S_iD_1}, \frac{m_{S_iD_1}(2^{R_t}-1)}{\bar{\gamma}}\Big)}{\Gamma(m_{S_iD_1})}\Bigg)\nonumber\nonumber\\
&\times\Bigg[\Bigg(1-\frac{\Gamma_{inc}\Big(m_{R_MD_1}, \frac{m_{R_MD_1}(2^{R_t}-1)}{\bar{\gamma}}\Big)}{\Gamma(m_{R_MD_1})}\Bigg)\nonumber\\
&+\sum_{j=1}^{M-1} \Bigg(1-\frac{\Gamma_{inc}\Big(m_{R_jR_{j+1}}, \frac{m_{R_jR_{j+1}}(2^{R_t}-1)}{\bar{\gamma}}\Big)}{\Gamma(m_{R_jR_{j+1}})}\Bigg)  \nonumber\\
&+\prod_{k=1}^{N} \Bigg(1-\frac{\Gamma_{inc}\Big(m_{S_kR_1}, \frac{m_{S_kR_1}(2^{R_t}-1)}{\bar{\gamma}}\Big)}{\Gamma(m_{S_kR_1})}\Bigg)\Bigg]
\end{align}

\end{theorem}

The outage expression (\ref{16}) is used for comparison with simulation results assuming the hypothesis of having the same value of shape factor $m$ on all links while the generalized outage expression (\ref{15}) is used to investigate the impact of different $m$ values over the different participating links on the overall end-to-end outage probability.\newline

The obtained outage probability is then compared with the simulated probability of decoding erroneous symbols at the receiver $D_1$ considering the equivalent instantaneous SNR between source node $S_1$ and destination node $D_1$. Many previous papers have been treating similar problems of outage probability computation for multiple relays cooperative or non cooperative networks. In NCC based 2 S-D pairs 2 relays wireless network, destination $D_1$ receives one signal from source node $S_2$ and another from the relay $R_2$ that has received the NC message from $R_1$ that combined its two received messages from $S_1$ and $S_2$. The general expression that shows all transmissions over all involved links with their respective attenuation effects and additive noises is given by
\begin{align}\label{25}
\tilde{y}_{D_1}=&\sqrt{P_{S_2}}x_{S_2}h_{S_2D1}+n_{S_2D_1}\nonumber\\
&+\sqrt{P_{R_2}}\Big[\sqrt{P_{R_1}}\Big(\sqrt{P_{S_1}}x_{S_1}h_{S_1R_1}+n_{S_1R_1}\nonumber\\
&+\sqrt{P_{S_2}}x_{S_2}h_{S_2R_1}+n_{S_2R_1}\Big)h_{R_1R_2}+n_{R_1R_2}\Big]\nonumber\\
&\times h_{R_2D_1} + n_{R_2D_1}
\end{align}
Expression (\ref{25}) is employed to determine the equivalent instantaneous SNR expression independently from the procedures of network coding at relay $R_1$ and decoding at destination $D_1$ that have no influence on the equivalent instantaneous receive SNR value.\newline
Hence, we derive the general expression of the global SNR illustrated in (\ref{18}).


\begin{figure}[b!]
\centerline{\includegraphics[scale=0.45]{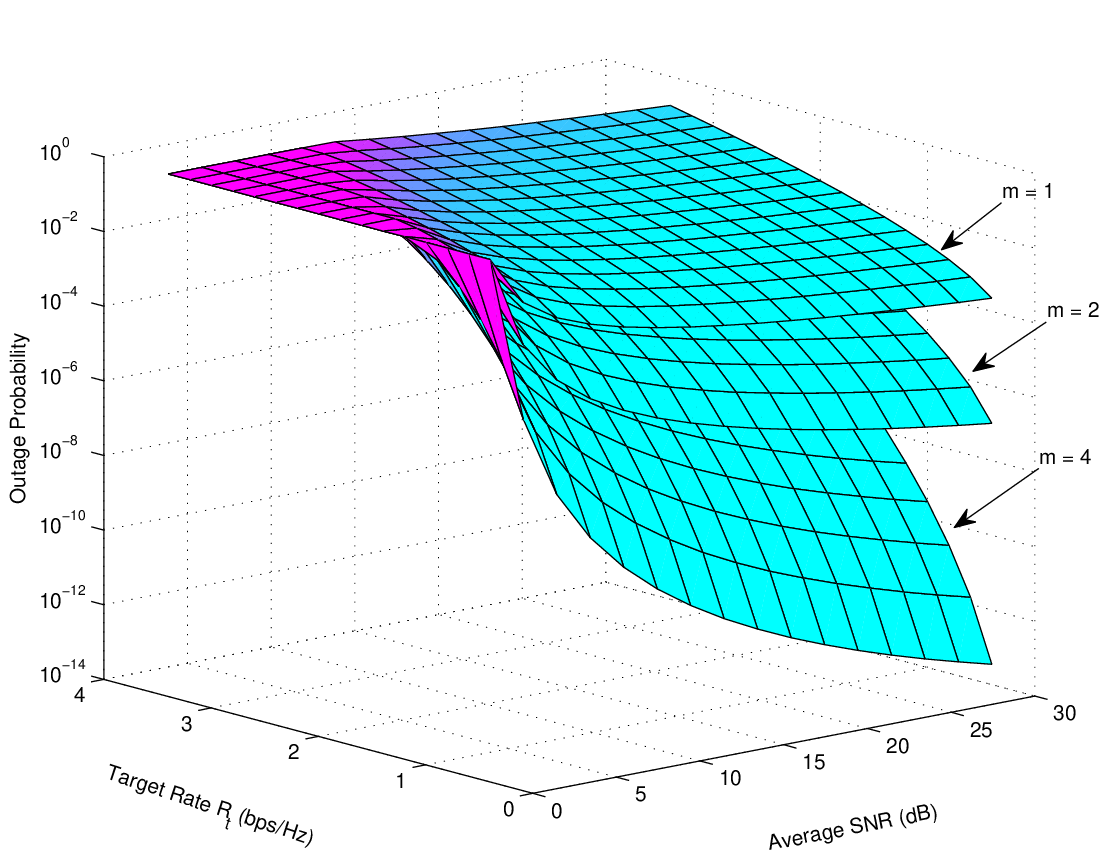}}
\caption{Outage probability of expression (\ref{16}) versus $R_t$ and average SNR for $m = 1, 2$ and $4$.}
\label{fig_6}
\end{figure}

A simpler version of (\ref{18}) can result from the following assumptions: independent and identically distributed fading channels with the same shape factor $m$ on all links, i.e. $h_{S_1R_1}=h_{S_2R_1}=h_{R_1R_2}=h_{R_2D_1}=h_{S_2D_1}=h$. Both sources $S_1$ and $S_2$ and both relays $R_1$ and $R_2$ have equal transmit powers $P_{S_1}=P_{S_2}=P_{R_1}=P_{R_2}=P$ and all added Gaussian noise variances have the same value $N_0$ at all receiving nodes, i.e. $E\{n_{S_2D_1}^2\}=E\{n_{R_2D_1}^2\}=E\{n_{R_1R_2}^2\}=E\{n_{S_1R_1}^2\}=E\{n_{S_2R_1}^2\}=N_0$. Therefore, (\ref{18}) yields
\begin{equation}\label{26}
\gamma_{S_1D1}^{eq} = \frac{P|h|^2\Big(1 + 2P^2|h|^4\Big)}{N_0\Big(2+P|h|^2+2P^2|h|^4\Big)}
\end{equation}

Simulated outage probability is the probability that the equivalent instantaneous SNR $\gamma_{S_1D1}^{eq}$ falls below average SNR threshold value $\gamma_{th}$, i.e.
\begin{align}\label{27}
P_{out}^{sim} &= P_r \Big\{\gamma_{S_1D1}^{eq} < \gamma_{th}\Big\}\nonumber\\
&=P_r\Bigg\{\frac{P|h|^2\Big(1 + 2P^2|h|^4\Big)}{N_0\Big(2+P|h|^2+2P^2|h|^4\Big)} < 2^{R_t}-1\Bigg\}
\end{align}

The simplified version of simulated outage probability given in (\ref{27}) is used to validate the analytic derived expression (\ref{16}) in the scenario of unique Nakagami fading factor $m$ in all links. However, for the most general case of different $m$ factors, expression (\ref{20}) of simulated outage probability serves to validate the generalized analytic expression given in (\ref{15}).

The diversity order deduced from outage probabilities of (\ref{17}) and (\ref{23}) in high SNR regions can be computed using the symbolic calculus of MATLAB.\newline
It is stated from \cite{24} that the diversity order $d$ can be expressed as:
\begin{equation}\label{28}
    d = - \lim_{SNR\to\infty}\frac{log(P_{out}(SNR))}{log(SNR)} 
\end{equation}
We use our derived outage probability to compute the diversity order using (\ref{28}) for different values of $m$ factor. For $m=1,~2,~3$ and $4$, we get the diversity values $d=2,~4,~6$ and $8$ respectively with expression (\ref{16}) of $P_{out}$ (two source-destination pairs) which leads to the result $d = 2m$ for NCC based wireless network. The result is the same for any value of target rate $R_t$ and any number of relays $M$.\newline
For an extension to $N$ ($S-D$) pairs, we use $P_{out}$ expression (\ref{22}) with different values of $N$ to confirm the following conclusion:\newline
The diversity order achieved in high SNR regions of the extended $N$ S-D pairs $M$ relays wireless network over Nakagami-$m$ fading channels, for any value of $R_t$ and any number of relays $M$, is given by 
\begin{equation}\label{29}
    d = Nm
\end{equation}
This result is validated only for integer values of fading parameter $m$.
The result of expression (\ref{29}) can be noticed, in case of $N=2$, on the graphs of outage probability for different values of $m$ in Fig. \ref{fig_4}. Actually, the diversity order $d$ is the absolute value of the slope of the outage probability at high SNRs.

\section{Simulation Results}
In this section, we use MATLAB software to perform SER simulations, analytic and simulated outage probability computation and finally, to determine a general reliable expression of the diversity order for the extended topology of wireless networks.
SER (or BER since we adopt binary symbols transmission) is obtained from the average of $10^3$ realizations in each one $10^4$ symbols are randomly generated at each source node along with $10^4$ Nakagami-$m$ fading coefficients for each link among the active ones involved in the end-to-end ($S_1-D_1$) pair communication. The received signal $\tilde{s}_1$ at $D_1$ is then compared to the original symbol generated at $S_1$. Figure \ref{fig_3} illustrates the results of BER simulations along with the theoretical outage probability computed with expression (\ref{16}). The perfect match of simulated BER for $m = 0.5,~1$ and $2$ with analytic outage probability confirms the exactness of our derived closed-form expression for a given value of target rate (spectral efficiency) $R_t$.\newline
In Fig. \ref{fig_4}, the curves of simulated outage probability of (\ref{27}) are almost completely fitting those of the analytic relation which confirms the high reliability of our derived closed-form expression. In order to evaluate the effect of Nakagami fading factor $m$ over the different links on the outage probability, we present some results on Fig. \ref{fig_5}. From these results, it can be seen that links ($S_1 \rightarrow R_1$) and ($S_2 \rightarrow R_1$) are not sensible to the variation of $m$. We notice almost the same conclusion for links ($R_1 \rightarrow R_2$) and ($R_2 \rightarrow D_1$). However, the deviation of $m$ value on link ($S_2 \rightarrow D_1$) has an important effect on the outage probability performance. These results can be crucial in processes like relay selection and power allocation for a better optimization of network resources management. A generalization for a wider range of SNR values and target rates $R_t$ is illustrated by the evolution of the outage probability for $m = 1,~2$ and $4$ in Fig. \ref{fig_6} where it is shown how increasing SNR allows to achieve higher spectral efficiency $R_t$.\newline

\section{Conclusion}
In this paper, an accurate closed-form expression of the theoretical outage probability was derived for the network coding and cooperation based 2 S-D pairs 2 relays wireless network over Nakagami-$m$ fading channels. We also presented a generalization of the analytic outage probability expression for an extended version of the considered wireless networks with $M$ relays and $N$ S-D pairs. The perfect match of the simulated outage probability evolution using the overall equivalent SNR confirmed the accuracy and reliability of our derived expressions which can serve to get the precise outage behaviour for such wireless networks and hence contribute in the network and resource optimization. Interesting results were also obtained regarding the role of Nakagami model fading parameter $m$ on different links in the outage probability behavior, which can be exploited in network configuration procedures like relay selection and power allocation. Moreover, we established a reliable expression of the diversity order. In a future work, the outage probability for the same wireless networks with application of error correction codes on all links will be investigated.

\balance

\end{document}